\documentclass[aps,preprint,showpacs]{revtex4}
\usepackage{amsfonts}
\usepackage{amsmath}
\usepackage{amssymb}
\usepackage{epsfig}

\renewcommand{\Re}{\mathop\mathrm{Re}\nolimits}

\begin{document}

\preprint{}
\title{Striking properties of Superconductor/Ferromagnet structures with
spin-dependent scattering}
\author{M. Faur\'{e}$^{1}$, A. I. Buzdin$^{1,2}$, A. A. Golubov$^{3}$, M.
Yu. Kupriyanov$^{4}$}
\affiliation{$^{1}$Centre de Physique Mol\'{e}culaire Optique et Hertzienne, Universit%
\'{e} Bordeaux 1-UMR 5798, CNRS, F-33405 Talence Cedex, France}
\affiliation{$^{2}$ Institut Universitaire de France, Paris, France}
\affiliation{$^{3}$Faculty of Science and Technology, University of Twente, 7500 AE
Enschede, The Netherlands}
\affiliation{$^{4}$Nuclear Physics Institute, Moscow State University, Moscow, 119992
Russia}

\begin{abstract}
We investigate Superconductor/Ferromagnet (S/F) hybrid structures in the
dirty limit, described by the Usadel equations. More precisely, the
oscillations of the critical temperature and critical current with the
thickness of the ferromagnetic layers are studied. We show that spin-flip
and spin-orbit scattering lead to the decrease of the decay length and the
increase of the oscillations period. The critical current decay is more
sensitive to these pair-breaking mechanisms than that of the critical
temperature. These two scattering mechanisms should be taken into account to
get a better agreement between experimental results and theoretical
description. We also study the influence of the interface transparency on
the properties of S/F structures.
\end{abstract}

\pacs{74.50.+r, 74.80.Dm, 75.30.Et}
\maketitle

\section{Introduction}

It has been known for quite a long time that superconductivity (S) and
ferromagnetism (F) are two antagonistic orderings and that they can hardly
coexist in a same compound. Although they tend to avoid each other, their
mutual interaction may be studied when they are spatially separated, which
is realized in Superconductor/Ferromagnet (S/F) hybrid structures (as a
review, see \cite{buzdinrev,golubov,bverev}). Indeed, in such systems,
superconductivity and ferromagnetism can influence each other through the so
called 'proximity effect'. The main peculiarity of the proximity effect in
S/F structures is the damping oscillatory behavior of the superconducting
correlations in the F layers (while they monotonically decay in normal
layers of Superconductor/Normal metal structures). In the dirty limit and
for large exchange field, the characteristic lengths of the decay and
oscillations are the same. They are given by $\xi _{f}=\sqrt{D_{f}/h}$,
where $D_{f}$ is the diffusion coefficient in the ferromagnet and $h$ is the
exchange field acting on the electrons spins. This unusual proximity effect
leads to several striking phenomena, such as the non-monotonic dependence of
the critical temperature and current of S/F multilayers on the F layers
thickness and the realization of the so-called $\pi $ junction in S/F/S
trilayers \cite{buzdinrev,golubov,bverev}.

Although the existing theory provides rather good qualitative description of
the observed\textit{\ }effects, there is still no complete quantitative
agreement with experiments. This indicates that besides the exchange field,
some additional pair-breaking mechanisms are present in the F layers.
Indeed, spin-flip process is inherent to the ferromagnetic layers (because
of magnetic impurities, spin-wave or non stoichiometric lattices) and may
have dramatic consequences on superconductivity (contrary to non magnetic
impurities that have very little impact). Such a pair-breaking mechanism
also arises in usually used weak ferromagnetic alloys, because they are
close to ferromagnetism disappearance and then quite favorable to large
magnetic disorder. This can be inferred for instance from the very strong
decrease of the critical current of S/F/S junctions as a function of the
thickness of the ferromagnetic layer in experimental studies \cite%
{ryazanov2001,sellier,ryazanov1}. In such experiments, the ferromagnetic
alloys used were Cu$_{x}$Ni$_{1-x}$ with $x\sim 0.5$, limit range of
concentration for ferromagnetic properties. In addition, the pair
destruction due to spin-orbit interaction must be taken into account as
well.

Though the spin-flip and spin-orbit interactions in the ferromagnetic
material were introduced in several recent theoretical studies on proximity
effect in S/F hybrids, only few simple limiting cases were considered (we'll
give appropriate references below). Thus the problem of quantitative
description of these effects in S/F multilayered systems is still unsolved.
In this paper we present the results of detailed theoretical study of the
influence of spin-flip and spin-orbit scattering mechanisms on the critical
temperature ($T_{c}^{\ast }$) and critical current ($I_{c}$) of S/F
multilayered systems. We obtain analytical and numerical solutions of the
problem which provide the basis not only for qualitative understanding of
experimental results but also to fit the data quantitatively. In particular,
we show that spin-orbit and spin-flip scattering mechanisms influence
differently the properties of S/F structures: spin-orbit mechanism can
destroy the $T_{c}^{\ast }$ and $I_{c}$ oscillations while spin-flip
scattering can only modify them. We also report on a striking non-intuitive
behavior that the critical temperature and current can acquire with the
variation of the S/F interface transparency\textit{.}

\section{Linearized Usadel equations}

A very convenient set of equations for an inhomogeneous superconductor was
elaborated by Eilenberger \cite{eilenberger}. However, the Eilenberger
equations can be replaced by the much simpler Usadel equations \cite{usadel}
when the electron scattering free path in S/F systems is shorter than the
superconducting length, which is often the case. These equations are
nonlinear but can be simplified when the temperature is close to the
critical temperature $T_{c}$ or at any temperature in the F layer when the
transparency is low. We consider a S/F multilayered system where all
physical quantities depend only on the coordinate $x$ perpendicular to the
layers. The natural choice of the spin-quantization axis is along the
direction of the exchange field. In the general case, magnetic and
spin-orbit scatterings mix up the up and down spin states. Therefore, two
anomalous Green functions, namely $F_{+}\sim \left\langle \psi _{\uparrow
}\psi _{\downarrow }\right\rangle $ and $F_{-}\sim \left\langle \psi
_{\downarrow }\psi _{\uparrow }\right\rangle $ are needed to describe this
situation. The linearized Usadel equations may be written for $\omega >0$ as
\cite{buzdinrev,houzet}
\begin{equation}
\left( \omega -\frac{D_{s}}{2}\frac{\partial ^{2}}{\partial x^{2}}\right)
F_{\pm s}\left( \omega ,x\right) =\Delta \left( x\right) ,
\label{Usadel_lin_S}
\end{equation}%
in the S layers and
\begin{equation}
\left( \omega -\frac{D_{f}}{2}\frac{\partial ^{2}}{\partial x^{2}}\pm ih+%
\frac{1}{\tau _{z}}+\frac{2}{\tau _{x}}\right) F_{\pm f}\left( \omega
,x\right) +\left( \frac{1}{\tau _{so}}-\frac{1}{\tau _{x}}\right) \left(
F_{\pm f}\left( \omega ,x\right) -F_{\mp f}\left( \omega ,x\right) \right)
=0,  \label{Usadel_lin_F}
\end{equation}%
in the F layers, where $\Delta \left( x\right) $ is the superconducting
order parameter, $h$ is the exchange field, $D_{s}\left( D_{f}\right) $ is
the diffusion coefficient in the S(F) layer and $\omega $ are the Matsubara
frequencies, $\omega =2\pi T\left( n+\frac{1}{2}\right) $. The parameter $%
\tau _{so}$ is the spin-orbit scattering time. The magnetic scattering times
are $\tau _{z}=\tau _{2}S^{2}/\left\langle S_{z}^{2}\right\rangle $ and $%
\tau _{x}=\tau _{2}S^{2}/\left\langle S_{x}^{2}\right\rangle $. The rate $%
\tau _{2}^{-1}$ is proportional to the square of the exchange interaction
potential (the notations are the same as in \cite{maki}). Note that the
microscopical Green functions theory of superconductors with magnetic
impurities and spin-orbit scattering was proposed by Abrikosov and Gor'kov
\cite{abrikosov}.

Besides, the usually used ferromagnets present a strong uniaxial anisotropy.
In that case, the perpendicular fluctuations of the exchange field are
suppressed, that is $\tau_{x}^{-1}\sim0$. Therefore, henceforth, $\tau_{z} $
will be noted as the magnetic scattering time $\tau_{m}$. The spin-flip
scattering is now simply incorporated replacing $\omega$ by $\omega+\frac {1%
}{\tau_{m}}$ in the standard Usadel equation, and $\tau_{m}$ is the magnetic
scattering time (see for example \cite{buzdin1985}). Note that the Usadel
equations in the F layers are not coupled anymore when $\tau_{so}^{-1}=0$,
and only one equation is needed.

\section{Theoretical description of the non-monotonic dependence of $%
T_{c}^{\ast}$.}

The common feature of all the S/F bilayered and multilayered
heterostructures is the non monotonic evolution of the critical temperature $%
T_{c}^{\ast }$ with the thickness of the ferromagnetic layer. This behavior
was first predicted by Buzdin and Kuprianov\textit{\ }\cite{kuprianov1} and%
\textit{\ }Radovic\textit{\ et al.} \cite{radovic}, and was since then
intensively studied both theoretically and experimentally (as a review, see
\cite{buzdinrev}). The presence of magnetic scattering can result in an
additional decrease of the transition temperature (on the contrary,
nonmagnetic impurities do not affect the transition temperature). Note that
the theoretical description of $T_{c}^{\ast }$ for ferromagnetic layers with
spin-orbit scattering was performed by Demler \textit{et al.} \cite{demler}
and Oh \textit{et al. }\cite{oh}. In the present section, we therefore
mainly focus on the influence of spin-flip process and neglect spin-orbit
scattering. We report here on the influence of the spin-flip scattering on
the non monotonic dependence of $T_{c}^{\ast }$ when the thickness of the
superconducting layer is supposed to be small,\textit{\ i. e. }$d_{s}\ll \xi
_{s}$. In that case, an analytical solution may be obtained. It should also
be underlined that the question of the spin-flip role was first addressed by
Tagirov \cite{tagirov} in the discussion of $T_{c}^{\ast }$ of SF
multilayers and by Fal'ko \textit{et al.} \cite{Falko} in the study of the
resistance of a diffusive F/N junction.

\subsection{Influence of spin-flip scattering on T$_{c}^{\ast}$}

We consider a S/F multilayered system with F layers of thickness $d_{f}$ and
parallel magnetization directions and S layers of thickness $d_{s}$, see
Fig. 1 (this system is also equivalent to a S/F bilayer of thicknesses $%
d_{f}/2$ and $d_{s}/2$ respectively). Further, we assume that the SF
interfaces are not 'magnetically-active', i.e. there is no rotation of the
quasipartice spin at the interfaces as considered in Refs. \cite%
{rainer,fogel1,fogel2,eschrig}. Under these conditions, long-range
spin-triplet superconductivity does not appear \cite{bverev}. Discussion of
the role of spin-flip scattering in a ferromagnet in combination with
magnetically-active SF interfaces and/or noncollinear magnetizations
requires separate study.

The anomalous Green's function $F_{s}$ varies a little in the S layer and
may be approximated by a simple expansion up to the second order $x^{2}$
(see for example \cite{vedyayev} and \cite{baladie})
\begin{equation}
F_{s}(x,\omega )=F_{0}\left( 1-\frac{\beta _{\omega }}{2}x^{2}\right) ,
\label{exp F}
\end{equation}%
where $F_{0}$ is the value of the anomalous Green's function at the center
of the S layer. Moreover, in that case, the spatial variation of the pair
potential $\Delta (x)$ can be neglected $\Delta (x)\sim \Delta $. It follows
from Eq. (\ref{Usadel_lin_S}) that $F_{0}=\frac{\Delta }{\omega +\tau
_{s}^{-1}}$ where $\tau _{s}^{-1}=\frac{D_{s}}{2}\beta _{\omega }$ is the
complex pair-breaking parameter.

\begin{figure}[tbp]
\includegraphics[width=2.8in ]{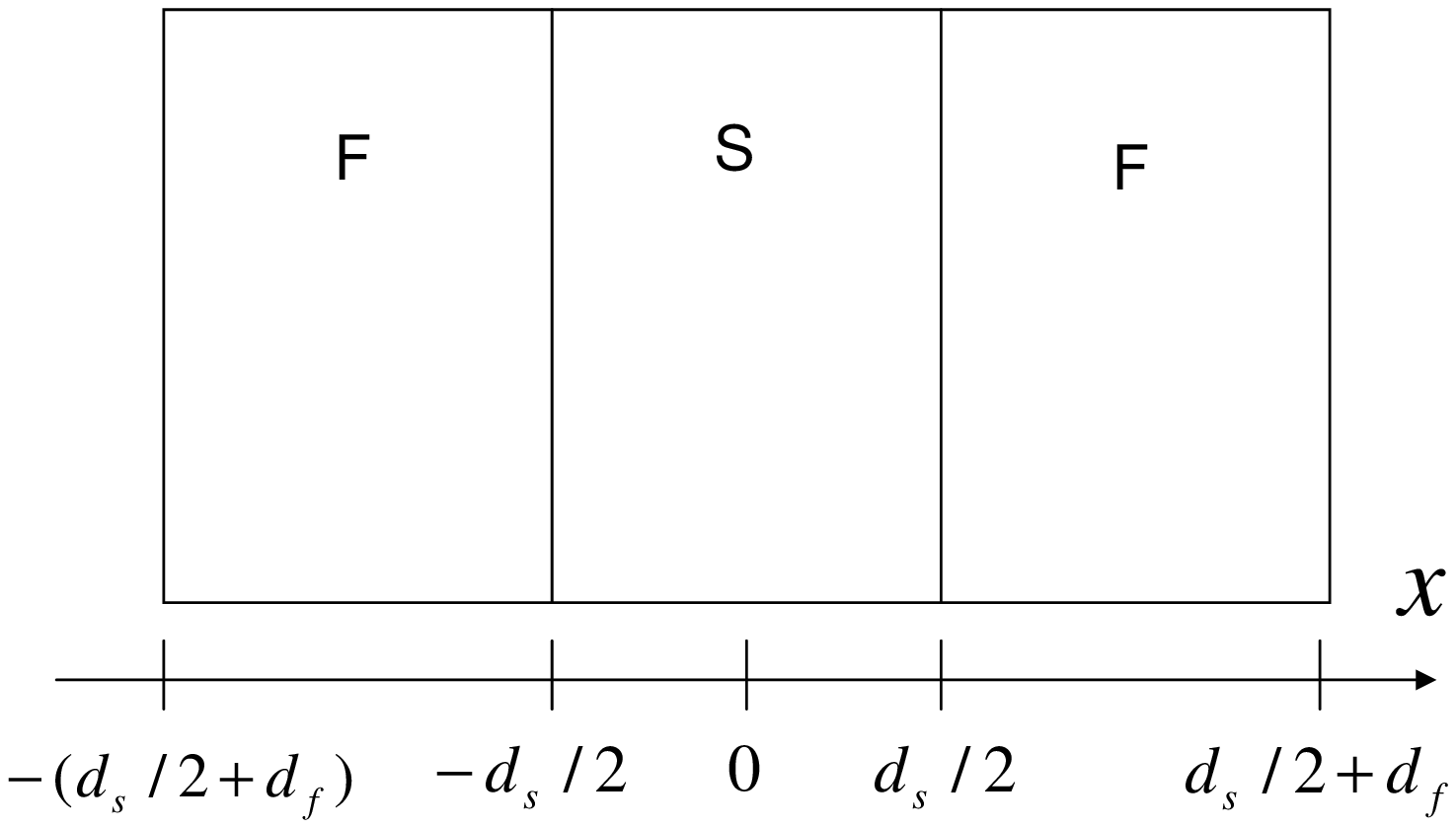}
\caption{Geometry of the considered system.}
\end{figure}

The Usadel equation in the S layer is completed by the self-consistency
equation%
\begin{equation}
\Delta(x)\ln\frac{T_{c}}{T_{c}^{\ast}}+\pi T_{c}^{\ast}\sum_{\omega}\left(
\frac{\Delta(x)}{\left\vert \omega\right\vert }-F_{\pm s}(x,\omega)\right)
=0,
\end{equation}
where $T_{c}$ is the bare transition temperature of the superconducting
layer in the absence of the proximity effect. Hence, this self-consistent
equation gives for $T_{c}^{\ast}$%
\begin{equation}
\ln\frac{T_{c}^{\ast}}{T_{c}}=\Psi\left( \frac{1}{2}\right) -\Re \Psi\left\{
\frac{1}{2}+\frac{1}{2\pi T_{c}^{\ast}\tau_{s}}\right\} .
\label{temperature critique}
\end{equation}
If the temperature variation is small ($\tau_{s}^{-1}\ll T_{c}$), (\ref%
{temperature critique}) becomes
\begin{equation}
\frac{T_{c}-T_{c}^{\ast}}{T_{c}}=\frac{\pi}{4T_{c}}\Re\left(
\tau_{s}^{-1}\right) .  \label{tau}
\end{equation}

The boundary conditions for the linearized Usadel equation are \cite%
{lukichev}
\begin{equation*}
\left( \frac{\partial F_{s}}{\partial x}\right) _{\pm d_{s}/2}=\frac {%
\sigma_{n}}{\sigma_{s}}\left( \frac{\partial F_{f}}{\partial x}\right) _{\pm
d_{s}/2},
\end{equation*}%
\begin{align}
F_{s}\left( d_{s}/2\right) & =F_{f}\left( d_{s}/2\right)
-\xi_{n}\gamma_{B}\left( \frac{\partial F_{f}}{\partial x}\right) _{d_{s}/2},
\\
F_{s}\left( -d_{s}/2\right) & =F_{f}\left( -d_{s}/2\right) +\xi
_{n}\gamma_{B}\left( \frac{\partial F_{f}}{\partial x}\right) _{-d_{s}/2}
\end{align}
with $\sigma_{n}\left( \sigma_{s}\right) $ the conductivity of the F(S)
layer, $\xi_{n}=\sqrt{\frac{D_{f}}{2\pi T_{c}}}$ and $\gamma_{B}=\frac {%
R_{b}\sigma_{n}}{\xi_{n}}$ is the interface transparency, related to the S/F
resistance per unit area $R_{b}$.They lead to the determination of the
pair-breaking parameter $\tau_{s}^{-1}$
\begin{equation}
\tau_{s}^{-1}=-\frac{D_{s}}{d_{s}}\frac{\sigma_{n}}{\sigma_{s}}\frac {%
F_{f}^{^{\prime}}(d_{s}/2)/F_{f}(d_{s}/2)}{1-\xi_{n}\gamma_{B}F_{f}^{^{%
\prime}}(d_{s}/2)/F_{f}(d_{s}/2)}.
\end{equation}

Next, the resolution of the Usadel equation in the F layers (\ref%
{Usadel_lin_F}) with symmetry consideration gives rise to the expression of
the anomalous Green's function in the F layer
\begin{equation}
F_{f}(x,\omega>0)=A\cosh\left[ k\left( x-d_{s}/2-d_{f}/2\right) \right] ,
\label{F0phase}
\end{equation}
in the 0 phase and
\begin{equation}
F_{f}(x,\omega>0)=B\sinh\left[ k\left( x-d_{s}/2-d_{f}/2\right) \right] ,
\end{equation}
in the $\pi$ phase. If $T_{c}<\tau_{m}^{-1}$, $h$, we may neglect the
Matsubara frequencies in the expression for $k$ that becomes
\begin{equation}
k=\frac{\sqrt{2}}{\xi_{f}}\sqrt{i+\alpha}=\frac{1}{\xi_{f1}}+i\frac{1}{%
\xi_{f2}},
\end{equation}
with $\xi_{f}=\sqrt{\frac{D_{f}}{h}}$ and $\alpha=\frac{1}{\tau_{m}h}$.%
\textit{\ }The two parameters $\xi_{f1}$ and $\xi_{f2}$ are respectively the
decay characteristic length and the oscillations period, and may be written
as%
\begin{align}
\xi_{f1} & =\frac{\xi_{f}}{\sqrt{\sqrt{1+\alpha^{2}}+\alpha}},  \label{decay}
\\
\xi_{f2} & =\frac{\xi_{f}}{\sqrt{\sqrt{1+\alpha^{2}}-\alpha}}.
\label{period}
\end{align}
For $\tau_{m}^{-1}=0$, Eqs. (\ref{decay}) and (\ref{period}) reduce to $%
\xi_{f1}=\xi_{f2}=\xi_{f}$. As expected, it is found that the decay length
and oscillations period are the same in absence of spin-flip scattering.

The pair breaking parameter may be determined and it reads%
\begin{equation}
\tau_{s,0}^{-1}(\omega>0)=\tau_{0}^{-1}\frac{q\tanh\left( q\widetilde{d_{f}}%
/2\right) }{1+\widetilde{\gamma}q\tanh\left( q\widetilde{d_{f}}/2\right) },
\label{tau-0}
\end{equation}
in the 0 phase and
\begin{equation}
\tau_{s,\pi}^{-1}(\omega>0)=\tau_{0}^{-1}\frac{q\coth\left( q\widetilde {%
d_{f}}/2\right) }{1+\widetilde{\gamma}q\coth\left( q\widetilde{d_{f}}%
/2\right) },  \label{tau Pi}
\end{equation}
in the $\pi$ phase, where $\tau_{0}^{-1}=\frac{D_{s}}{d_{s}}\frac{\sigma_{n}%
}{\sigma_{s}}\frac{1}{\xi_{f}}$, $\widetilde{\gamma}=\frac{\xi_{n}}{\xi_{f}}%
\gamma_{B}$, $q=k\xi_{f}$ and $\widetilde{d_{f}}=\frac{d_{f}}{\xi_{f}}$.

\bigskip

First, if the interface is supposed to be transparent, analytical
expressions of the variation of the temperature may be found if $\frac{%
T_{c}-T_{c}^{\ast}}{T_{c}}\ll1$:
\begin{align}
\frac{4\tau_{0}}{\pi}\left( T_{c}-T_{c}^{\ast0}\right) & =\frac{1}{2}\frac{%
a\sinh(a\widetilde{d_{f}})-b\sin(a\widetilde{d_{f}})}{\cosh ^{2}(a\widetilde{%
d_{f}})\cos^{2}(b\widetilde{d_{f}})+\sin^{2}(b\widetilde {d_{f}})\sinh^{2}(a%
\widetilde{d_{f}})}, \\
& \\
\frac{4\tau_{0}}{\pi}\left( T_{c}-T_{c}^{\ast\pi}\right) & =\frac{1}{2}\frac{%
a\sinh(a\widetilde{d_{f}})+b\sin(b\widetilde{d_{f}})}{\cosh ^{2}(a\widetilde{%
d_{f}})\sin^{2}(b\widetilde{d_{f}})+\cos^{2}(b\widetilde {d_{f}})\sinh^{2}(a%
\widetilde{d_{f}})},
\end{align}
where two dimensionless parameters have been introduced, namely $a=\frac {%
\xi_{f}}{\xi_{f1}}=\sqrt{\sqrt{1+\alpha^{2}}+\alpha}$ and $b=\frac{\xi_{f}}{%
\xi_{f2}}=\sqrt{\sqrt{1+\alpha^{2}}-\alpha}$, so that $q=a+ib$.

In the general case however a numerical analysis has to be performed.

The ratio of the characteristic lengths\textit{\ }%
\begin{equation}
\frac{\xi_{f1}}{\xi_{f2}}=\frac{\sqrt{\sqrt{1+\alpha^{2}}-\alpha}}{\sqrt {%
\sqrt{1+\alpha^{2}}+\alpha}},
\end{equation}
clearly shows that the magnetic scattering decreases the decay length and
increases the oscillation period. If $\tau_{m}^{-1}>>h$, $\xi_{f1}$ can
become much smaller than $\xi_{f2}$. Besides, the decrease of $\xi_{f1}$
makes the observation of the oscillations more difficult.

The evolution of the critical temperature without and with spin-flip
scattering is given in Fig. 2. The phase that really occurs is the one with
higher critical temperature. It is seen that $\xi_{f1}$ decreases in
presence of spin-flip while $\xi_{f2}$ increases.

\begin{figure}[tbp]
\includegraphics[width=2.8in ]{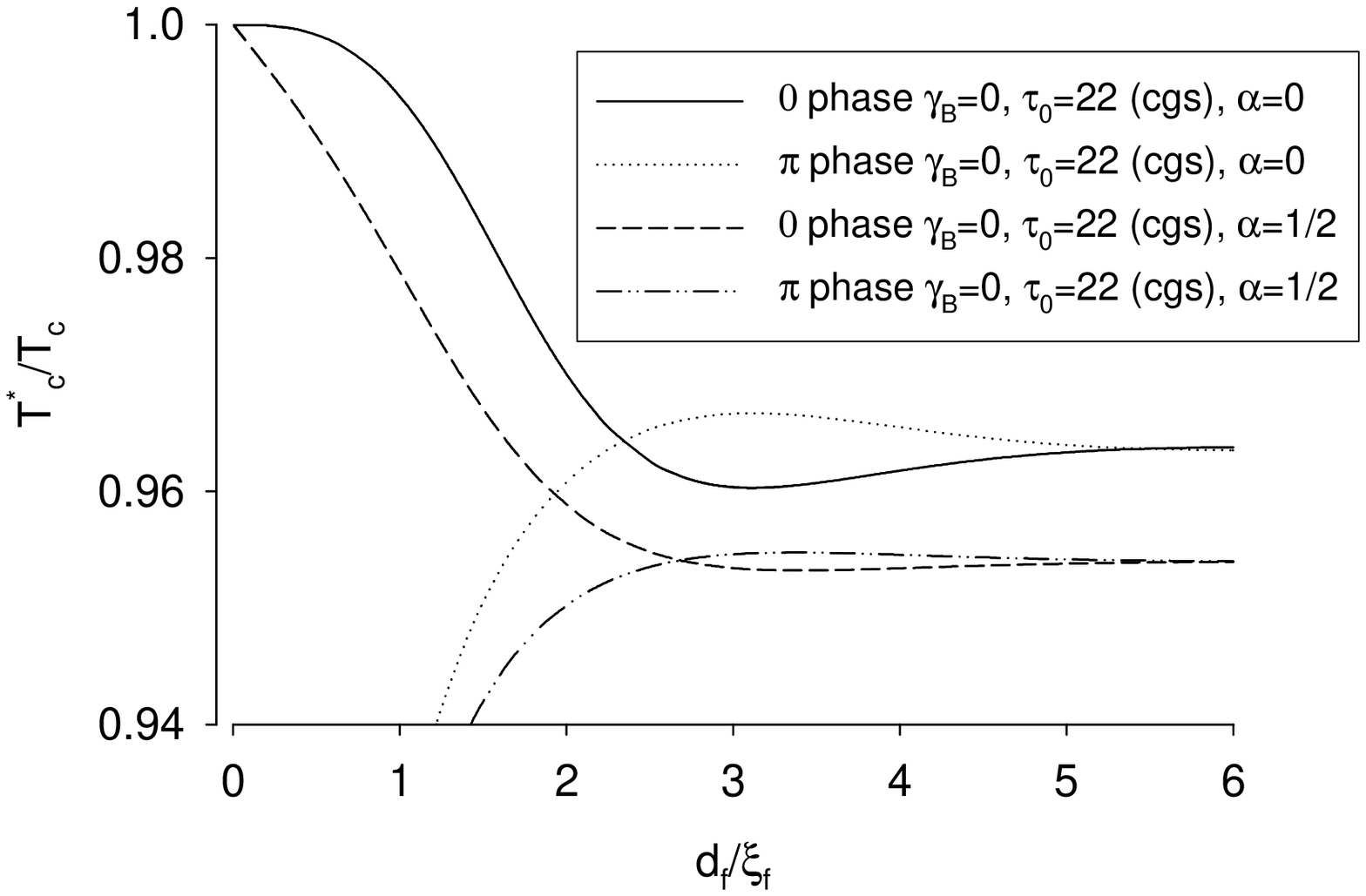}
\caption{Influence of the spin-flip scattering on the evolution of critical
temperature as a function of the ferromagnetic layer thickness.}
\end{figure}

\subsection{Influence of the interface transparency parameter $\protect%
\gamma _{B}$ on $T_{c}^{\ast }$}

The influence of the interface transparency parameter $\gamma _{B}$ on $%
T_{c}^{\ast }$ of SF bilayers was studied before by Aarts \textit{et al.}
\cite{Aarts}, Fominov \textit{et al.} \cite{Fominov} and Tagirov \cite%
{tagirov}. Here we extend this discussion taking into account spin-flip
scattering in the F-layer and considering broader range of interface\
transparencies and demonstrate that new effects take place in this situation.

The intriguing evolution of $T_{c}^{\ast 0}$ with the interface transparency
parameter $\widetilde{\gamma }$ must be underlined (see Fig. 3). If $\alpha
=0$, there is no magnetic scattering and it could intuitively be believed
that, the higher the barrier, the less is the influence of the proximity
effect on the S layer and therefore, $T_{c}^{\ast }\left( \widetilde{\gamma }%
\gg 1\right) >T_{c}^{\ast }\left( \widetilde{\gamma }\sim 1\right) $.
However, it can be seen from Fig. 3 that the critical temperature is a
decreasing function of the interface transparency parameter $\widetilde{%
\gamma }$ for a small thickness of the F layer. This counter-intuitive
behavior can be qualitatively understood for a S/F bilayer. The probability
for a Cooper pair to leave the S layer is smaller for a low transparent
interface $\left( \widetilde{\gamma }\gg 1\right) $. Nevertheless, the
probability for this pair to come back again in the S layer is much higher
for a transparent interface. Indeed, when the F layer is thin enough, the
reflection of the Cooper pair at the other interface of the F layer allows
the pair to cross again the first interface, which is easier when $%
\widetilde{\gamma }$ is small. Consequently, the staying time in the F layer
increases with the increase of the barrier, and when this time becomes
higher than the coherence time of the Cooper pair, the pair is destroyed,
leading to a weakened superconductivity. Therefore, the critical temperature
decreases with the barrier in that case. On the other hand, if $\widetilde{%
d_{f}}$ increases, the Cooper pair is hardly reflected by the external
interface of the F layer whatever the value of $\widetilde{\gamma }$ is and
the critical temperature is expected to increase with the barrier.

\begin{figure}[tbp]
\includegraphics[width=2.8in ]{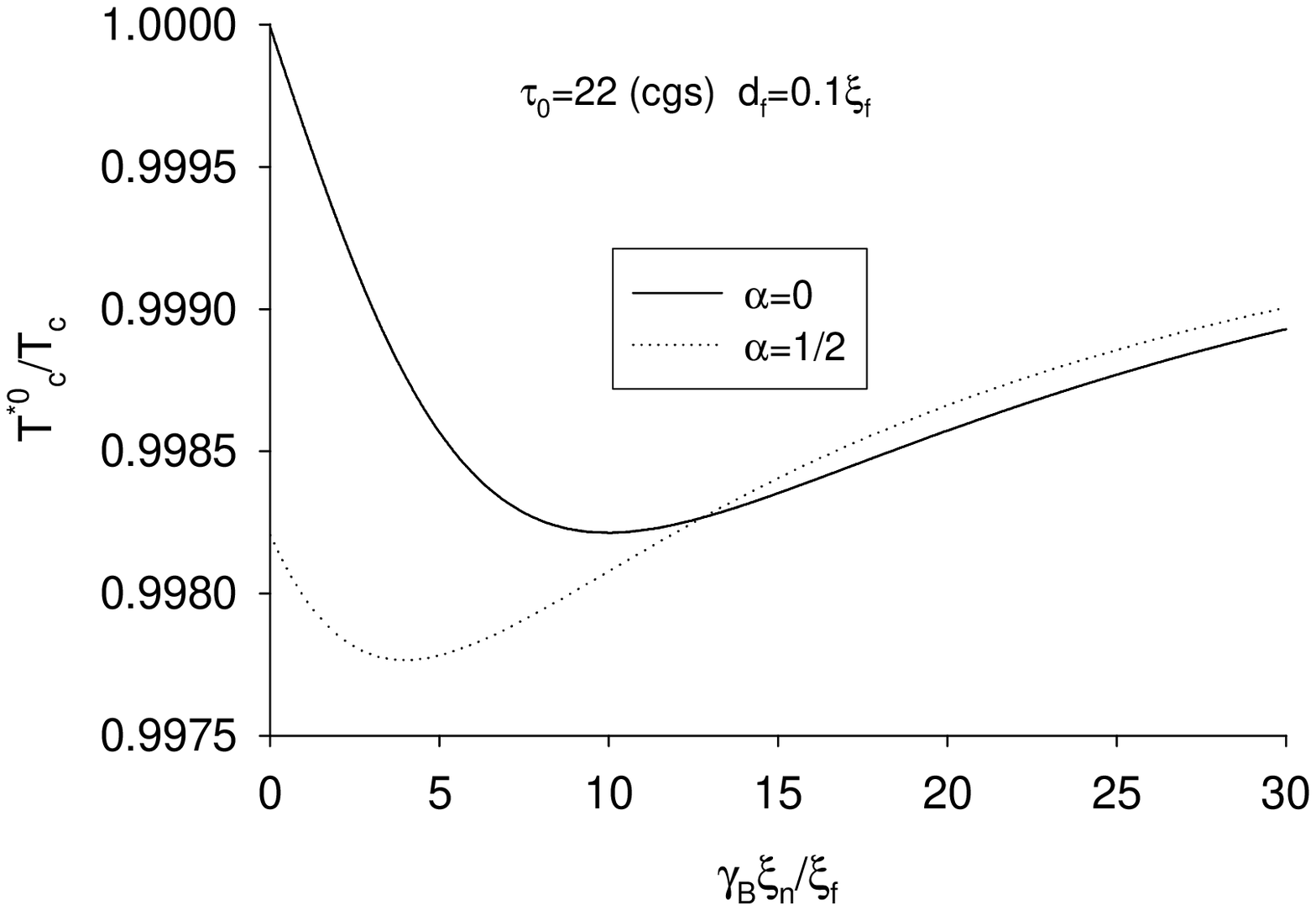}
\caption{Evolution of the critical temperature versus the interface
transparency parameter for a S/F bilayer.}
\end{figure}

Besides, an analytical approach using formula (\ref{tau}) leads to the
determination of the critical temperature when $\widetilde{d_{f}}<1$. In
that case, the pair breaking parameter in the 0 phase becomes
\begin{equation}
\Re\left( \tau_{m}^{-1}\right) =\tau_{0}^{-1}\frac {\widetilde{\gamma}+\frac{%
\widetilde{d_{f}}}{6}}{\left( \widetilde{\gamma }+\frac{\widetilde{d_{f}}}{6}%
\right) ^{2}+\left( \frac{1}{\widetilde{d_{f}}}\right) ^{2}}.
\end{equation}
Two behaviors emerge whether $\widetilde{\gamma}\widetilde{d_{f}}>>1$ or
not. Indeed, if $\widetilde{\gamma}>\widetilde{d_{f}}^{-1}$, the critical
temperature evolution is described by
\begin{equation}
\left( \frac{T_{c}^{\ast}}{T_{c}}\right) _{\widetilde{\gamma}\widetilde {%
d_{f}}>1}=1-\frac{\pi}{4T_{c}}\frac{\tau_{0}^{-1}}{\widetilde{\gamma}},
\end{equation}
which is an increasing function of $\widetilde{\gamma}$. On the contrary, if
$\widetilde{\gamma}<\widetilde{d_{f}}^{-1}$, the critical temperature
decreases when $\widetilde{\gamma}$ increases, following
\begin{equation}
\left( \frac{T_{c}^{\ast}}{T_{c}}\right) _{\widetilde{\gamma}\widetilde {%
d_{f}}<1}=1-\frac{\pi}{4T_{c}}\widetilde{d_{f}}^{2}\tau_{0}^{-1}\widetilde{%
\gamma}.  \label{Tc1}
\end{equation}
The critical temperature evolution with $\widetilde{\gamma}$ presented in
Fig. 3 is therefore understood.

In presence of small magnetic scattering, the critical temperature evolution
remains qualitatively the same. As shown in Fig. 3, it appears that
superconductivity may be less destroyed when there is spin-flip scattering.
In that case, the pair breaking parameter becomes
\begin{equation}
\Re\left( \tau_{m}^{-1}\right) =\tau_{0}^{-1}\frac {\widetilde{\gamma}+\frac{%
\alpha}{\widetilde{d_{f}}}+\frac{\widetilde{d_{f}}}{6}}{\left( \widetilde{%
\gamma}+\frac{\alpha}{\widetilde{d_{f}}}+\frac{\widetilde{d_{f}}}{6}\right)
^{2}+\left( \frac{1}{\widetilde{d_{f}}}\right) ^{2}}.
\end{equation}
The critical temperature is described by
\begin{equation}
\left( \frac{T_{c}^{\ast}}{T_{c}}\right) _{\widetilde{\gamma}\widetilde {%
d_{f}}>1}=\left( \frac{T_{c}^{\ast}\left( \alpha=0\right) }{T_{c}}\right) _{%
\widetilde{\gamma}\widetilde{d_{f}}>1}+\frac{\pi}{4T_{c}}\frac{\tau_{0}^{-1}%
}{\widetilde{\gamma}}\frac{\alpha}{\widetilde{d_{f}}\widetilde{\gamma}},
\end{equation}
when $\widetilde{d_{f}}\widetilde{\gamma}>1$, and magnetic scattering leads
to a slight enhancement of the transition temperature. On the contrary, if $%
\widetilde{\gamma}\widetilde{d_{f}}<1,$
\begin{equation}
\left( \frac{T_{c}^{\ast}}{T_{c}}\right) _{\widetilde{\gamma}\widetilde {%
d_{f}}<1}=\left( \frac{T_{c}^{\ast}\left( \alpha=0\right) }{T_{c}}\right) _{%
\widetilde{\gamma}\widetilde{d_{f}}<1}-\frac{\pi}{T_{c}}\widetilde{d_{f}}%
\tau_{0}^{-1}\alpha,
\end{equation}
and spin-flip implies the decrease of the transition temperature.

\subsection{Influence of spin-orbit scattering and 'perpendicular' spin-flip}

Let us now consider briefly the general case, with spin-orbit and/or
perpendicular spin-flip scattering. An additional parameter has to be
introduced, namely
\begin{equation}
\alpha_{\perp}=\frac{1}{h}\left( \frac{1}{\tau_{x}}-\frac{1}{\tau_{so}}%
\right) ,
\end{equation}
and the parameter $\alpha$ now becomes
\begin{equation}
\alpha=\frac{1}{h}\left( \frac{1}{\tau_{z}}+\frac{2}{\tau_{x}}\right) .
\end{equation}
In that case, expressions (\ref{tau-0}) and(\ref{tau Pi}) are modified. In
the 0 phase, $q\tanh\left( q\widetilde{d_{f}}\right) $ is replaced by
\begin{equation}
q\tanh\left( q\widetilde{d_{f}}/2\right) +\beta\frac{q^{\ast}\tanh\left(
q^{\ast}\widetilde{d_{f}}/2\right) -q\tanh\left( q\widetilde{d_{f}}/2\right)
}{\beta+\frac{1+\widetilde{\gamma}q^{\ast}\tanh\left( q^{\ast }\widetilde{%
d_{f}}/2\right) }{1+\widetilde{\gamma}q\tanh\left( q\widetilde{d_{f}}%
/2\right) }},
\end{equation}
where $q$ becomes
\begin{equation}
q^{2}=2\left( \frac{\omega}{h}+\alpha-\alpha_{\bot}+i\sqrt{1-\alpha_{\bot
}^{2}}\right) ,
\end{equation}
and $\beta$ is
\begin{equation}
\beta=-\alpha_{\perp}\frac{\alpha_{\perp}-i\sqrt{1-\alpha_{\perp}^{2}}}{1+%
\sqrt{1-\alpha_{\perp}^{2}}}.
\end{equation}
In the $\pi$ phase, $q\coth\left( q\widetilde{d_{f}}/2\right) $ is replaced
by
\begin{equation}
q\coth\left( q\widetilde{d_{f}}/2\right) +\beta\frac{q^{\ast}\coth\left(
q^{\ast}\widetilde{d_{f}}/2\right) -q\coth\left( q\widetilde{d_{f}}/2\right)
}{\beta+\frac{1+\widetilde{\gamma}q^{\ast}\coth\left( q^{\ast }\widetilde{%
d_{f}}/2\right) }{1+\widetilde{\gamma}q\coth\left( q\widetilde{d_{f}}%
/2\right) }}.
\end{equation}
Therefore, the influence of 'perpendicular' spin-flip scattering and
spin-orbit scattering is quite similar to the influence of 'parallel'
spin-flip processes, in the sense that it also implies the decrease of the
decaying length and the increase of the oscillations period. However, a
special situation arises when $\alpha_{\perp}>1$. Then, the oscillations of
the Cooper pair wave function are completely destroyed. Similar conclusion
for spin-orbit mechanism was obtained in \cite{demler}. In fact, the
influence of the 'perpendicular' magnetic scattering is analogous to the
spin-orbit scattering (see Fig. 4). Probably the role of 'perpendicular'
spin-flip or spin-orbit scattering is important for the understanding of
experimental results where no oscillation of the critical temperature was
detected. Besides, note that the critical temperature oscillations can not
disappear when there is only 'parallel' spin-flip.

\begin{figure}[tbp]
\includegraphics[width=2.8in ]{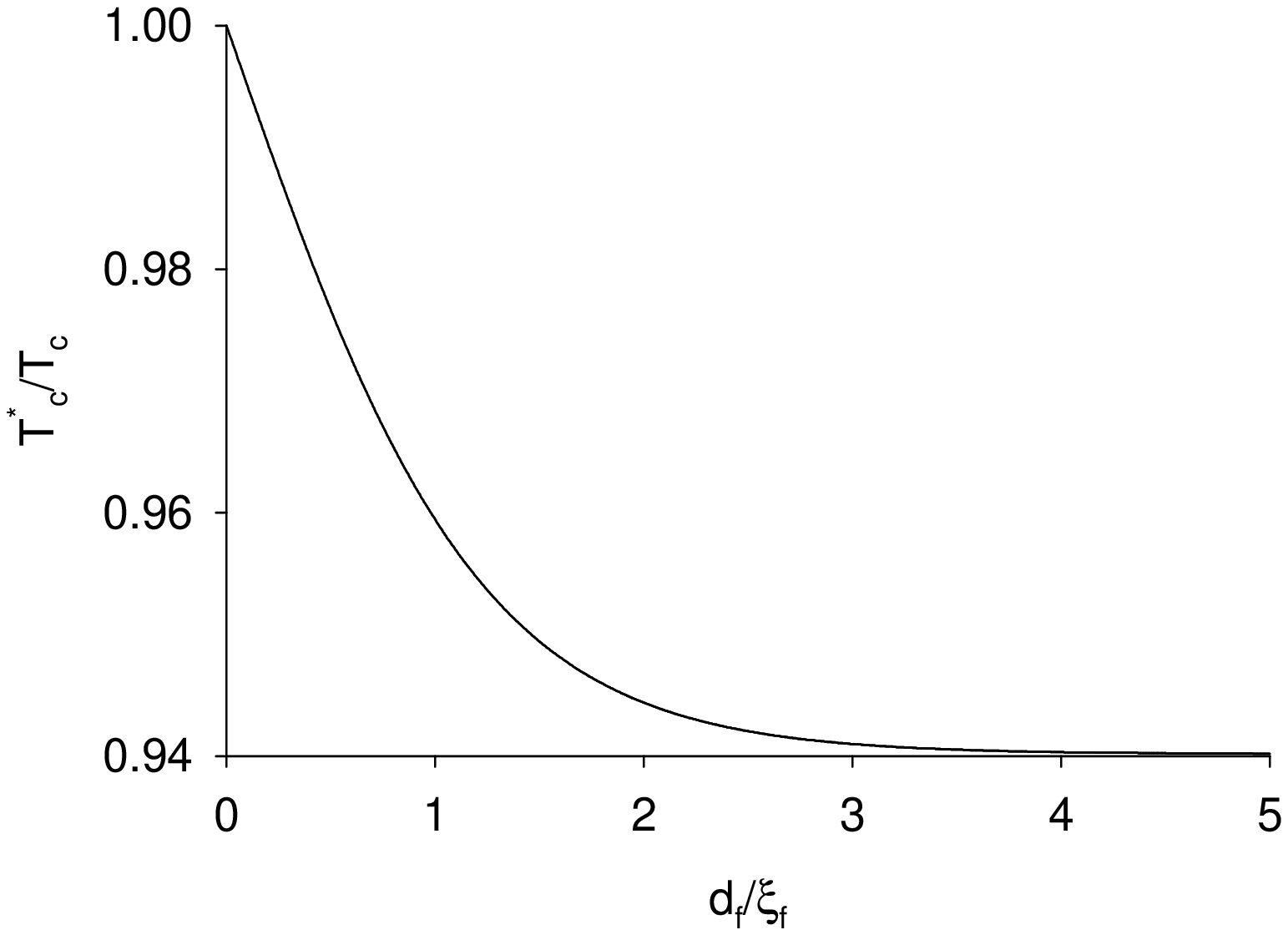}
\caption{Monotonic evolution of the critical temperature for an transparent
interface, and for $\protect\tau_{0}=22$ (cgs), $\protect\alpha=1$ and $%
\protect\alpha_{\perp}=-3/2$.}
\end{figure}

\section{Behavior of the critical current I$_{c}$}

A Josephson junction may be realized with S/F/S sandwiches in which the weak
link between the two superconductors is ensured by the ferromagnetic layer.
The supercurrent $J_{s}\left( \varphi\right) $ flowing across the structure
can be expressed as $J_{s}\left( \varphi\right) =I_{c}\sin\left(
\varphi\right) $, where $I_{c}$ is the critical current and $\varphi$ stands
for the phase difference between the two superconducting layers. A standard
junction has at equilibrium $I_{c}>0$ and $\varphi=0$, and therefore, no
current exists. It may appear however that $I_{c}$ becomes negative, which
implies that the equilibrium phase difference is $\varphi=\pi$ and the
ground state undergoes a $\pi$ phase shift. This so called '$\pi$ junction'
was first predicted for S/F/S structures in the clean limit \cite{buzdin1982}%
, and later in the more realistic case of the diffusive limit \cite%
{kuprianov}. The critical current $I_{c}$ in S/F/S junctions is controlled
by the exchange field in the ferromagnet, the interface transparency
parameter $\gamma_{B}$, the spin-flip and spin-orbit scattering rates. In
this section we will discuss the influence of these parameters on $I_{c}$,
paying the main attention to the role of spin-orbit and magnetic
scatterings. Besides, note that the spin-orbit effect on the critical
current has already been studied for S/F bilayers separated by an insulator
with thin F layers \cite{krivoruchko} and for the FSFSF structure with
noncollinear magnetizations of F-layers \cite{bve}. We will consider a
symmetric S/F/S structure with the F layer having thickness $d_{f}$ and
assume that the dirty limit conditions are fulfilled for S and F materials.
We will assume that the rigid boundary condition $\gamma_{B}>>\min\{%
\sigma_{n}\xi_{S}/\sigma_{s}\xi_{n},1\}$ is fulfilled, when the suppression
of superconductivity in S can be neglected and the linearized Usadel
equations (\ref{Usadel_lin_F}) can be used. The case of transparent
interfaces, $\gamma_{B}=0$, will be considered at the end.

\subsection{General expression for I$_{c}$}

To derive the general expression for $I_{c}$ we should supplement (\ref%
{Usadel_lin_F}) by the boundary conditions \cite{lukichev} at S/F interfaces
$(x=\pm d_{f})$%
\begin{equation}
\gamma_{B}\xi_{n}\frac{\partial}{\partial x}F_{\pm f}(\pm d_{f})=\pm G_{S}%
\left[ \frac{\Delta(\pm d_{f})}{\omega}-F_{\pm f}(\pm d_{f})\right]
sign(\omega),  \label{BC_SFS}
\end{equation}%
\begin{equation*}
G_{S}=\frac{\omega}{\sqrt{\omega^{2}+\left\vert \Delta\right\vert ^{2}}}%
,\quad\Delta(\pm d_{f})=\left\vert \Delta\right\vert \exp\left\{ \pm
i\varphi/2\right\} .
\end{equation*}
The solution of the boundary problem (\ref{Usadel_lin_F}), (\ref{BC_SFS})
has the form
\begin{equation}
F_{+f}=\frac{1}{2}((1+\eta_{\omega})\left\{ A_{\omega+}\cosh\left[ k_{+}x%
\right] +B_{\omega+}\sinh\left[ k_{+}x\right] \right\}
+(1-\eta_{\omega})\left\{ A_{\omega-}\cosh\left[ k_{-}x\right] +B_{\omega
-}\sinh\left[ k_{-}x\right] \right\} ),  \label{SolP}
\end{equation}%
\begin{equation}
F_{-f}=\frac{1}{2}((1-\eta_{\omega})\left\{ A_{\omega+}\cosh\left[ k_{+}x%
\right] +B_{\omega+}\sinh\left[ k_{+}x\right] \right\}
-(1+\eta_{\omega})\left\{ A_{\omega-}\cosh\left[ k_{-}x\right] +B_{\omega
-}\sinh\left[ k_{-}x\right] \right\} ),  \label{SolM}
\end{equation}
where the coefficients $A_{\omega\pm}$ and are given $B_{\omega\pm}$ in the
Appendix, while
\begin{equation}
\eta_{\omega}=\frac{\alpha_{so}+\sqrt{\alpha_{so}^{2}-1}}{ihsign(\omega )}%
,\quad k_{\pm}=\frac{\sqrt{2}}{\xi_{f}}\sqrt{\left\vert \widetilde{\omega }%
\right\vert +\alpha_{so}+\alpha\pm\sqrt{\alpha_{so}^{2}-1}}.  \label{EtiBe}
\end{equation}
The parameter $\alpha_{so}$ is defined by $\alpha_{so}=1/(h\tau_{so})$, $%
\alpha=1/(h\tau_{m})$ and $\widetilde{\omega}=\omega/h$. Note that in the
general case 'perpendicular' spin-flip has to be added in $\alpha_{so}$ and $%
\alpha$. The general expression for the supercurrent is

\begin{equation}
J_{S}=\frac{i\pi T\sigma_{n}}{4e}\sum_{\omega=-\infty,\sigma=\pm}^{\infty }%
\left[ \tilde{F}_{\sigma f}\frac{\partial}{\partial x}F_{\sigma f}-F_{\sigma
f}\frac{\partial}{\partial x}\tilde{F}_{\sigma f}\right] ,  \label{curr}
\end{equation}
where $\tilde{F}_{\pm f}(x,\omega)=F_{\pm f}^{\ast}(x,-\omega)$ .
Substituting (\ref{SolP}) and (\ref{SolM}) into the above expression and
taking into account the symmetry relations (given in the Appendix), we get $%
J_{S}=I_{c}\sin\varphi$.

For large spin-orbit scattering, i. e. $\tau_{so}^{-1}\geq h$, the critical
current $I_{c}$ equals to
\begin{equation}
\frac{e\xi_{n}I_{c}}{2\pi T_{c}\sigma_{n}}=\frac{T}{T_{c}}\sum_{\omega
=0}^{\infty}\frac{\Delta^{2}G_{S}^{2}}{\omega^{2}(1+\eta_{\omega}^{2})}%
(\kappa_{+}+\eta_{\omega}^{2}\kappa_{-}),  \label{Ics}
\end{equation}
where
\begin{equation}
\kappa_{\pm}=\frac{\xi_{n}k_{\pm}}{(\gamma_{B}^{2}\xi_{n}^{2}k_{%
\pm}^{2}+G_{S}^{2})\sinh\left[ k_{\pm}d_{f}\right] +2G_{S}\gamma_{B}%
\xi_{n}k_{\pm }\cosh\left[ k_{\pm}d_{f}\right] }.
\end{equation}

At larger exchange field $h\geq\tau_{so}^{-1}$, the critical current becomes
\begin{equation}
\frac{eI_{c}}{2\pi T_{c}\sigma_{n}}=\frac{T}{\xi_{f}T_{c}}\sum_{\omega
=0}^{\infty}\frac{\Delta^{2}G_{S}^{2}}{\omega^{2}}\Re\left\{ \frac{(a+ib)%
\left[ 1+i\frac{2}{\sqrt{\alpha_{so}^{-2}-1}}\right] }{((\gamma_{B}%
\xi_{n}k)^{2}+G_{S}^{2})\sinh\left[ kd_{f}\right] +2G_{S}\gamma_{B}\xi_{n}k%
\cosh\left[ kd_{f}\right] }\right\} ,  \label{Ic}
\end{equation}
where $k=q/\xi_{f}$ and $q=a+ib$, with
\begin{equation}
a=\sqrt{\left\vert \widetilde{\omega}\right\vert +\alpha_{so}+\alpha +\sqrt{%
(\left\vert \widetilde{\omega}\right\vert +\alpha)^{2}+2(\left\vert
\widetilde{\omega}\right\vert +\alpha)\alpha_{so}+1}},  \label{p}
\end{equation}%
\begin{equation}
b=\sqrt{\frac{1-\alpha_{so}^{2}}{\left\vert \widetilde{\omega}\right\vert
+\alpha_{so}+\alpha+\sqrt{(\left\vert \widetilde{\omega}\right\vert
+\alpha)^{2}+2(\left\vert \widetilde{\omega}\right\vert +\alpha)\alpha_{so}+1%
}}}.  \label{q}
\end{equation}

In appropriate limits, expressions (\ref{Ics}),(\ref{Ic}) transform into the
previously obtained results \cite{buzdinrev}, \cite{demler}, \cite{lukichev},%
\cite{baladie},\cite{buzdin2003}. This approach is valid if $F_{\omega \pm
}\ll 1$ or
\begin{equation}
\gamma _{B}\gg \gamma ,~\frac{a}{a^{2}+b^{2}}\frac{\Delta }{\pi T_{c}}\frac{1%
}{\min \left\{ 1,d_{f}/\xi _{n}\right\} }.  \label{Cond_gamma_B}
\end{equation}%
\qquad

As follows from (\ref{Ic}) - (\ref{q}), in the case of strong\ spin-orbit
scattering $\tau_{so}^{-1}\geq h,$ the critical current decays monotonically
with the increase of $d_{f}$ with two decay lengths $k_{\pm}^{-1}$ defined
by Eq.(\ref{EtiBe}). It is seen from (\ref{EtiBe}) that in the limit $%
h\rightarrow0$, the parameter $\eta_{\omega}\rightarrow\infty.$ As a result,
the contribution to the critical current in (\ref{Ics}) comes only from $%
\kappa_{-}$ component with the length scale $k_{-}^{-1}$ which describes the
case of an S/N/S junction in which spin-orbit scattering does not influence $%
I_{c}.$ With the increase of $h$, the contribution to $I_{c}$ from the
faster decaying $\kappa_{+}$ component $(k_{+}>k_{-})$ also increases and
the difference between $k_{+}$ and $k_{-}$ decreases$.$ Finally, when $%
h=\tau _{so}^{-1}$, both scales coincide, $k_{+}=k_{-}$, and the components $%
\kappa_{+},\kappa_{-}$ provide equal contributions to $I_{c}.$

For relatively weak spin-orbit scattering, $\tau_{so}^{-1}\leq h$, the
dependence $I_{c}(d_{f})$ follows the damped oscillation law, when two
length scales $\xi_{f1},\xi_{f2}$ can be introduced describing respectively
the decay and the oscillation period of $I_{c}(d_{f})$. In this section, we
will concentrate on the case $\tau_{so}^{-1}\leq h$. The scales $\xi_{f1}$, $%
\xi_{f2}$ are related to $a,b$ and will be discussed in detail below in
different limits.

\subsubsection{Limit of small F layer thickness and large $\protect\gamma%
_{B}.$}

In the limit of small\ thickness $d_{f}\ll \xi _{f1},$ and large interface
transparency parameter $\gamma _{B}\gg d_{f}/\xi _{n},(\pi T_{c}G_{S}\xi
_{n})/(d_{f}(\omega +\tau _{m}^{-1}+\tau _{s0}^{-1})$ we can neglect the
terms $G_{S}^{2}$ and $G_{S}\gamma _{B}\xi _{n}k(kd_{f})^{2}$ in the
denominator of Eq.(\ref{Ic}) and with the accuracy of better than $%
(d_{f}/\xi _{f1})^{3}$ get
\begin{equation}
\frac{eR_{N}I_{c}}{4\pi T_{c}}=\frac{T}{\widetilde{\gamma }d_{f}T_{c}}\xi
_{f}\sum_{\omega =0}^{\infty }\frac{\Delta ^{2}G_{S}^{2}}{\omega ^{2}}%
\left\{ \frac{\Omega _{1}+2\alpha _{so}+(\Omega ^{2}-v^{2}+4\alpha
_{so}\Omega )\frac{d_{f}{}^{2}}{6\xi _{f}^{2}}}{\Omega _{1}^{2}+v^{2}+\frac{%
d_{f}{}^{2}}{3\xi _{f}^{2}}\Omega (\Omega ^{2}+v^{2})}\right\} ,
\label{Ic_small_d}
\end{equation}%
where $R_{N}=2R_{B}$ is the normal junction resistance, $\widetilde{\gamma }%
=\gamma _{B}\xi _{n}/\xi _{f},$ $\widetilde{\omega }=\omega /h$ and%
\begin{equation*}
\Omega =\widetilde{\omega }+\alpha _{so}+\alpha ,\quad v^{2}=1-\alpha
_{so}^{2},\quad \Omega _{1}=\Omega +2G_{S}\frac{\pi T_{c}\xi _{n}}{\gamma
_{B}hd_{f}}.
\end{equation*}%
If additionally
\begin{equation*}
\gamma _{B}\gg \frac{6\xi _{n}^{3}\pi T_{c}}{d_{f}{}^{3}\left[ (\pi
T_{c}/h+\alpha _{so}+\alpha )^{2}+1-\alpha _{so}^{2})\right] h}
\end{equation*}%
then
\begin{equation}
\frac{eR_{N}I_{c}}{4\pi T_{c}}=\frac{\xi _{f}}{\widetilde{\gamma }d_{f}}%
\left\{ \frac{T}{T_{c}}\sum_{\omega =0}^{\infty }\frac{\Delta ^{2}}{\omega
^{2}+\Delta ^{2}}\frac{\Omega +2\alpha _{so}}{(\Omega ^{2}+v^{2})}-\frac{%
d_{f}{}^{2}}{6\xi _{f}^{2}}\frac{\Delta }{2T_{c}}\tanh \frac{\Delta }{2T}%
\right\} .  \label{Ic_small_da}
\end{equation}%
For $h\gg \pi T_{c},\tau _{m}^{-1},\tau _{s0}^{-1}$ and $T\ll T_{c}$, the
sum in (\ref{Ic_small_da}) can be calculated by transforming from summation
into integration over $\omega $ resulting in\textit{\ }
\begin{equation}
\frac{eR_{N}I_{c}}{4\pi T_{c}}=\frac{\Delta }{2T_{c}}\frac{\xi _{f}}{%
\widetilde{\gamma }d_{f}}\left[ \frac{2\Delta }{h\pi }\ln \frac{h}{\Delta }+%
\frac{3}{h\tau _{so}}+\frac{1}{h\tau _{m}}-\frac{d_{f}{}^{2}}{6\xi _{f}^{2}}%
\right] ,  \label{Ic_small_da0}
\end{equation}%
From (\ref{Ic_small_da0}) it follows that the transformation from $0$ to $%
\pi $ junction may occur when the thickness $d_{f}$ exceeds some critical
value $d_{\pi }$
\begin{equation}
d_{f}\gtrsim d_{\pi },\quad d_{\pi }=\sqrt{6}\xi _{f}\sqrt{\frac{2\Delta }{%
\pi h}\ln \frac{h}{\Delta }+\frac{3}{h\tau _{so}}+\frac{1}{h\tau _{m}}}.
\label{cond_pi_sd}
\end{equation}%
For $\tau _{so}^{-1},\tau _{m}^{-1}\ll h$\ this result agrees with \cite%
{buzdin2003}. It is interesting to note that\textit{\ }in this range of
parameters\textit{,} the condition (\ref{cond_pi_sd}) of the transition to $%
\pi $ state does not depend on the properties of the interfaces. As is also
seen from Eq. (\ref{cond_pi_sd}), both spin-orbit and spin-flip scattering
increase the thickness $d_{\pi }$ corresponding to the first $0$ to $\pi $
transition.

If $\widetilde{\gamma }d_{f}/\xi _{f}\lesssim 1$,\ then the term $G_{S}\xi
_{f}/(\widetilde{\gamma }d_{f})$\ in (\ref{Ic_small_d}) is not small. As a
result, the transition to a $\pi $ state should depend on the properties of
interfaces and occurs at $h$ larger than the critical value $h_{\pi }\propto
\pi T_{c}(\xi _{n}/d_{f})$ following from Eq.(\ref{Ic_small_da0}).

Therefore, in the limit of small $d_{f}\ll \xi _{f1}$ and large $\gamma
_{B}\gg 1$ the transition from $0$ to $\pi $ junction exists only\ if the
exchange energy $h\gtrsim h_{\pi }\propto \pi T_{c}(\xi _{n}/d_{f})$
sufficiently exceeds $\pi T_{c}$. The smaller the interface transparency
parameter $\gamma _{B}$,\ the larger should be the exchange energy $h$. The
possibility of the 0-$\pi $ transition in junctions with small thickness $%
d_{f}$ in the case of large $\gamma _{B}$ is related to the multiple
scattering at the boundaries. As a result, the electrons spend more time in
the F layer and its influence on superconductivity is enhanced. This is the
manifestation of the same mechanism that leads in S/F bilayers to the
critical temperature decrease with the increase of the interface
transparency parameter (see section II B).

\subsubsection{Limit of large F layer thickness$.$}

In the limit of large $d_{f}\gg\xi_{f1}$ for the critical current from (\ref%
{Ic}) - (\ref{q}), we have
\begin{equation}
\frac{eI_{c}}{2\pi T_{c}\sigma_{n}}=\frac{2T}{\xi_{f}T_{c}}\sum_{\omega
=0}^{\infty}\frac{\Delta^{2}G_{S}^{2}}{\omega^{2}}\frac{u\sin(bd_{f}/\xi
_{f})+v\cos(bd_{f}/\xi_{f})}{\left( (G_{S}+a\widetilde{\gamma})^{2}+%
\widetilde{\gamma}^{2}b^{2}\right) ^{2}}\exp\left\{ -a\frac{d_{f}}{\xi_{f}}%
\right\} ,  \label{Ic_Ldg}
\end{equation}
where the coefficients $u$ and $v$ are defined by%
\begin{align}
u & =b\left( G_{S}^{2}-(a^{2}+b^{2})\widetilde{\gamma}^{2}\right) +\frac{2}{%
\sqrt{\alpha_{so}^{-2}-1}}G_{S}^{2}\left( a+\widetilde{\gamma }\left(
a^{2}+b^{2}\right) \left( 2G_{S}+a\widetilde{\gamma}\right) \right) ,
\label{uv} \\
v & =G_{S}^{2}a+\widetilde{\gamma}\left( a^{2}+b^{2}\right) \left( 2G_{S}+a%
\widetilde{\gamma}\right) -\frac{2}{\sqrt{\alpha_{so}^{-2}-1}}b\left(
G_{S}^{2}-(a^{2}+b^{2})\widetilde{\gamma}^{2}\right) .  \notag
\end{align}

If additionally $\min\left\{ h,\tau_{so}^{-1},\tau_{m}^{-1}\right\} \gg\pi
T_{c},$ then both $a$ and $b$ may be considered as independent on Matsubara
frequencies, since the sum in (\ref{Ic}) converges at $\omega\approx\pi
T_{c} $. In this case
\begin{equation}
\frac{e\xi_{f}I_{c}}{2\pi T_{c}\sigma_{n}}=\left[ \Sigma_{1}\sin\frac{d_{f}}{%
\xi_{f2}}+\Sigma_{2}\cos\frac{d_{f}}{\xi_{f2}}\right] \exp\left\{ -\frac{%
d_{f}}{\xi_{f1}}\right\} ,  \label{Id1}
\end{equation}
where
\begin{equation}
\Sigma_{1}=\frac{2T}{T_{c}}\sum_{\omega=0}^{\infty}\frac{\Delta^{2}G_{S}^{2}%
}{\omega^{2}}\frac{u}{\left( (G_{S}+a\widetilde{\gamma})^{2}+b^{2}\widetilde{%
\gamma}^{2}\right) ^{2}},  \label{S1}
\end{equation}%
\begin{equation}
\Sigma_{2}=\frac{2T}{T_{c}}\sum_{\omega=0}^{\infty}\frac{\Delta^{2}G_{S}^{2}%
}{\omega^{2}}\frac{v}{\left( (G_{S}+a\widetilde{\gamma})^{2}+b^{2}\widetilde{%
\gamma}^{2}\right) ^{2}}.  \label{S2}
\end{equation}
The two characteristic length scales are given by the following expressions
\begin{equation}
\xi_{f1}=\xi_{f}\sqrt{\frac{1}{\alpha+\alpha_{so}+\sqrt{\alpha^{2}+2\alpha%
\alpha_{so}+1}}},  \label{ksi11}
\end{equation}%
\begin{equation}
\xi_{f2}=\xi_{f}\sqrt{\frac{\alpha+\alpha_{so}+\sqrt{\alpha^{2}+2\alpha
\alpha_{so}+1}}{1-\alpha_{so}^{2}}},  \label{ksi21}
\end{equation}
which generalize Eqs. (\ref{decay}) and (\ref{period}) for the case of the
presence of the spin-orbit scattering in a ferromagnet. One can see that
with the increase of both scattering rates $\tau_{m}^{-1}$ and $%
\tau_{so}^{-1}$ the decay length $\xi_{f1}$ decreases, while the oscillation
period $\xi_{f2}$ increases.

For a weak exchange field $h<<\pi T_{c}$ and sufficiently high temperatures $%
\pi T>>\{\tau_{so}^{-1},\tau_{m}^{-1},h\}$, with $h>\tau_{so}^{-1}$, only
the first term with $n=0$ in (\ref{Ic}) is important and we have
\begin{equation}
\frac{e\xi_{f}I_{c}}{2\pi T_{c}\sigma_{n}}=\frac{2\Delta^{2}G_{0}}{\pi
^{2}TT_{c}}\frac{\Sigma_{1}\sin(d_{f}/\xi_{f2})+\Sigma_{2}\cos(d_{f}/\xi
_{f2})}{\left( (G_{0}+a_{0}\widetilde{\gamma})^{2}+b_{0}^{2}\widetilde {%
\gamma}^{2}\right) ^{2}}\exp\left\{ -\frac{d_{f}}{\xi_{f1}}\right\} .
\label{Id2}
\end{equation}
where $G_{0}=\pi T/\sqrt{(\pi T)^{2}+\Delta^{2}}$ and $\Sigma_{1}$ and $%
\Sigma_{2}$ become
\begin{equation}
\Sigma_{1}=\frac{2T}{T_{c}}\frac{\Delta^{2}G_{0}^{2}}{(\pi T)^{2}}\frac{u_{0}%
}{\left( (G_{0}+a_{0}\widetilde{\gamma})^{2}+b_{0}^{2}\widetilde{\gamma}%
^{2}\right) ^{2}},  \label{S11}
\end{equation}%
\begin{equation}
\Sigma_{2}=\frac{2T}{T_{c}}\frac{\Delta^{2}G_{0}^{2}}{(\pi T)^{2}}\frac{v_{0}%
}{\left( (G_{0}+a_{0}\widetilde{\gamma})^{2}+b_{0}^{2}\widetilde{\gamma}%
^{2}\right) ^{2}}.  \label{S21}
\end{equation}
The parameters $a_{0}$, $b_{0}$, $u_{0}$, $v_{0}$ are obtained replacing $%
\omega$ by $\pi T$ in expressions (\ref{p}), (\ref{q}), (\ref{uv}). The two
characteristic lengths may be written in that case as
\begin{equation}
\xi_{f1}=\xi_{f}\sqrt{\frac{1}{\pi T/h+\alpha+\alpha_{so}+\sqrt{(\pi
T/h+\alpha)^{2}+2(\pi T+\alpha)\alpha_{so}+1}}},  \label{ksi12}
\end{equation}%
\begin{equation}
\xi_{f2}=\xi_{f}\sqrt{\frac{\pi T/h+\alpha+\alpha_{so}+\sqrt{(\pi
T/h+\alpha)^{2}+2(\pi T+\alpha)\alpha_{so}+1}}{1-\alpha_{so}^{2}}.}
\label{ksi22}
\end{equation}
For $\tau_{so}^{-1},\tau_{m}^{-1}=0$,\ Eqs.(\ref{ksi12}), (\ref{ksi22})
reduce to a simple expression $\xi_{f1,2}^{-1}=\xi_{n}^{-1}[(h^{2}/(\pi
T_{c})^{2}+(T/T_{c})^{2})^{1/2}\pm(T/T_{c})]^{1/2}$\ \cite{ryazanov2001}
which describes the temperature variations of both length scales. More
precisely, in the considered limits, $\xi_{f1}^{-1}\sim\xi_{n}^{-1}\sqrt{%
2T/T_{c}}$, while $\xi_{f2}^{-1}\sim\xi_{f1}^{-1}h/(2\pi\sqrt{TT_{c}}%
)<<\xi_{f1}^{-1}$. One can see that the scattering rates $\tau_{so}^{-1}$,$%
\tau_{m}^{-1}$ make the $T$ variation of $\xi_{f1,2}$ weaker.

From (\ref{Id1}), (\ref{Id2}), one can derive the condition for $I_{c}=0$
and obtain
\begin{equation}
d_{fn}=d_{f},\quad\frac{d_{fn}}{\xi_{f2}}=\pi n-\arctan(\frac{\Sigma_{2}}{%
\Sigma_{1}}),\quad n=0,1,2,...  \label{zeros}
\end{equation}
when the transitions between $0-$ and $\pi$-states occur. The position of
the first zero, $d_{f1},$ depends both on the material parameters of the
ferromagnetic layer and the properties of the interfaces and superconducting
electrodes, while the distance between the zeros is the function only of $%
\xi_{f2}$ and therefore depends only on the transport parameters of the
ferromagnetic material.

From the structure of coefficients $u$ and $v$ (see Eq.(\ref{uv}) and (\ref%
{zeros})), it follows that in the limit of small $\gamma_{B}$ ($F_{\pm f}=%
\frac{\Delta}{\sqrt{\omega^{2}+\Delta^{2}}}\exp\left( \pm i\varphi /2\right)
$ at S/F interfaces) and\textit{\ }$\min\left\{
h,\tau_{so}^{-1},\tau_{m}^{-1}\right\} \gg$\textit{\ }$\pi T_{c}$%
\begin{equation}
\frac{d_{fk}}{\xi_{f1}}=\pi k-\arctan(\frac{2b-a\sqrt{\alpha_{so}^{-2}-1}}{%
2a+b\sqrt{\alpha_{so}^{-2}-1}}),\text{ }k=1,2,...  \label{zerSim}
\end{equation}

\bigskip In particular, it follows from (\ref{zerSim}) that for small
spin-orbit and spin-flip scatterings $h\gg\tau_{so}^{-1},\tau_{m}^{-1}$,
\begin{equation}
\frac{d_{fk}}{\xi_{f1}}\approx\pi k-\frac{\pi}{4},  \label{zerSS}
\end{equation}
and the well known result $d_{f1}\approx(3\pi/4)\xi_{f2}$\ for the first
critical thickness for 0 to $\pi$\ state is reproduced. This $d_{f1}$ value
approximately satisfies the condition of validity of the large $d_{f}$
approximation considered in this section.

It follows from (\ref{zerSim}) that the critical thickness $d_{f1}$\
increases\ with $\tau_{m}^{-1}$ and $\tau_{so}^{-1}$ (see also numerical
results below).

An increase of $\gamma _{B}$\ results in the suppression of the magnitude of
$F_{\pm f}(\pm d_{f})$\ $\sim \Delta (1+\gamma _{B})^{-1}$\ near the S/F
interfaces, which leads to the decrease of $d_{f1}.$\ Formally it is due to
the fact that with an increase of $\gamma _{B}$, the coefficient $b$\ in (%
\ref{uv}) and, hence, $\Sigma _{1},$\ may change its sign resulting in the
existence of the solution $d_{f0}$\ of (\ref{zeros}) for $n=0.$\ This
solution corresponds to the thickness range in which simple large $d_{f}$\
approximation (\ref{Ic_Ldg}) is no valid anymore. This fact is in the full
agreement with our consideration performed for the limit of small $d_{f}$.
Namely, it follows from (\ref{cond_pi_sd}) that at large $\gamma _{B}$\ and $%
h$\ we have%
\begin{equation}
d_{f1}=\sqrt{6}\xi _{f}\sqrt{\frac{2\Delta }{\pi h}\ln \frac{h}{\Delta }%
+\alpha +3\alpha _{so}}\ll \xi _{f1}.  \label{DF0_Lg}
\end{equation}

Note that in actual experimental situations, when the approximation $%
d_{f}\gg\xi_{f1}$ is not fulfilled, simple expressions (\ref{Id1}), (\ref%
{Id2}), (\ref{ksi11}), (\ref{ksi12}), (\ref{ksi21}),(\ref{ksi22}) are not
valid and to analyze the data it is convenient to introduce the effective
decay length $\xi_{f1}^{eff}$%
\begin{equation}
\xi_{f1}^{eff}=\frac{1}{d_{m1}-d_{m2}}\ln\frac{I_{cm1}}{I_{cm2}},
\label{ksi1_eff}
\end{equation}
where $d_{m1,2}$ are the thicknesses at which the first $(I_{cm2})$ and
second $(I_{cm2})$ maxima of $I_{c}(d_{f})$ occur.

Below we will focus on the influence of $\gamma_{B},$ $\tau_{so}^{-1},$ and$%
\ \tau_{m}^{-1}$ on the critical current.

\subsection{Influence of interface transparency parameter $\protect\gamma%
_{B} $ on I$_{c}$}

Consider first the simplest case of vanishing $\tau _{so}^{-1}$, $\tau
_{m}^{-1}$. In this limit, we immediately deduce from (\ref{Ic})-(\ref{q})
that
\begin{equation}
\frac{e\xi _{f}I_{c}}{2\pi T_{c}\sigma _{n}}=\frac{T}{T_{c}}\sum_{\omega
=0}^{\infty }\frac{\Delta ^{2}G_{S}^{2}}{\omega ^{2}}\Re\left\{ \frac{a+ib}{%
((\gamma _{B}\xi _{n}k)^{2}+G_{S}^{2})\sinh \left[ kd_{f}\right]
+2G_{S}\gamma _{B}\xi _{n}k\cosh \left[ kd_{f}\right] }\right\} ,
\label{IcT0}
\end{equation}%
where the coefficients $a$ and $b$ become
\begin{equation}
a=\sqrt{\left\vert \widetilde{\omega }\right\vert +\sqrt{\left\vert
\widetilde{\omega }\right\vert ^{2}+1}},  \label{pT0}
\end{equation}%
\begin{equation}
b=\sqrt{\left\vert \widetilde{\omega }\right\vert +\sqrt{\left\vert
\widetilde{\omega }\right\vert ^{2}+1}}.  \label{qT0}
\end{equation}%
The dependence of the critical current as a function of the F layer
thickness calculated from (\ref{IcT0}) for $T=0.5T_{c},$\ and $h=3\pi T_{c}$%
\ for different values of the interface transparency parameter $\gamma _{B}$%
\ is presented in Fig. 5. Note that for small $\gamma _{B}$ Eq.(\ref{IcT0})
is formally not applicable, since is was derived under the condition (\ref%
{Cond_gamma_B}) of sufficiently large $\gamma _{B}$ . Therefore, for $\gamma
_{B}=$0, $\gamma _{B}=$1 and $\gamma _{B}=$5 we have added for comparison
the corresponding curves calculated numerically by direct solution of the
Usadel equations for arbitrary $\gamma _{B}$, which show that Eq.(\ref{IcT0}%
) provides reasonable approximation even in the small $\gamma _{B}$ range.
It is clearly seen from Fig. 5 that with an increase of $\gamma _{B}$,\ the
position of the first zero $d_{f1}$\ is shifted into the region $d_{f}<\xi
_{f1}.$\ As discussed above, the increase of $\gamma _{B}$\ results in the
suppression of $F_{\pm f}(\pm d_{f})$\ in the F layer, such that $ReF_{\pm
f}(x)$\ changes sign in the F layer center at smaller $d_{f}$\ .

The results plotted in Fig. 5 make it possible to estimate the upper limit
for the $I_cR_N$ product of the SFS junction in a $\pi$-state. This upper
limit can be achieved in the case of highly transparent interfaces ($%
\gamma_B $=0) and in the absence of spin-flip and spin-orbit scattering. As
follows from Fig. 5, $I_cR_N^{max} \approx 0.1 \pi T_c$ that provides $%
I_cR_N^{max} \approx 250 \mu V $ in the case of Nb electrodes ($T_c$=9K).

\begin{figure}[tbp]
\includegraphics[width=2.8in ]{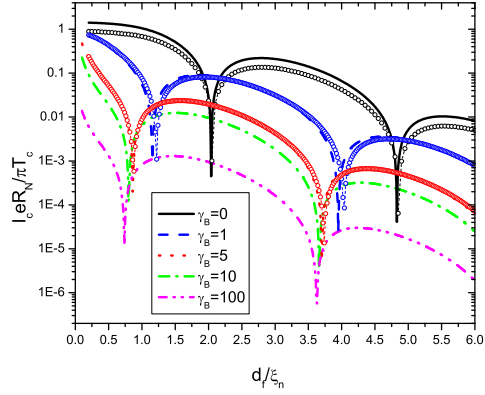}
\caption{(Color on line) Influence of the interface transparency parameter $%
\protect\gamma _{B}$ on the thickness dependence of the critical current in
S/F/S junction for $\protect\alpha =\protect\alpha _{so}=0$, $h=3\protect\pi %
T_{c}$ and $\frac{T}{T_{c}}=0.5$. Open circles: the results of exact
numerical calculations for $\protect\gamma _{B}=$0, $\protect\gamma _{B}=$1
and $\protect\gamma _{B}=$5 (from top to bottom).}
\end{figure}

At large F layer thickness, $I_{c}(d_{f})$\ is determined by Eqs.(\ref{Id1}%
), (\ref{Id2}). If spin-flip and spin-orbit scattering are negligible, the
ratio $\xi_{f1}/\xi_{f2}=h/(T+\sqrt{(\pi T)^{2}+h^{2}})$\ follows from (\ref%
{ksi12}), (\ref{ksi22}) and depends only on $h$\ and $T.$\ For typical
ferromagnets $h\gtrsim\pi T_{c}$\ and $\xi_{f1}$\ approximately equals to $%
\xi_{f2}.$\ However, if the spin-flip scattering and spin-orbit rates become
relatively large $\tau_{m}^{-1},\tau_{so}^{-1}\gtrsim h,$\ the situation may
change drastically. Consider first the influence of spin-flip scattering on I%
$_{c}.$

\subsection{Influence of spin-flip scattering on I$_{c}$}

If spin-flip scattering is not negligible, then the ratio of the
characteristic lengths in the decaying solution (\ref{Id1}), (\ref{Id2})
becomes
\begin{equation}
\frac{\xi_{f1}}{\xi_{f2}}=\frac{1}{\pi T/h+\alpha+\sqrt{(\pi T/h+\alpha
)^{2}+1}},  \label{R_sf}
\end{equation}
and for strong spin-flip scattering $\tau_{m}^{-1}\gtrsim h$ the decay
length $\xi_{f1}$ may become much smaller than the oscillation period $%
\xi_{f2}$.\ This results in the much stronger decrease of $I_{c}$ versus $%
d_{f}$ in S/F/S junctions.

The evolution of $I_{c}$\ for different values of $\tau_{m}^{-1}$\
calculated from (\ref{Ic}) for $T=0.5T_{c},$\ $h=3\pi T_{c}$\ and $%
\gamma_{B}=10$\ is given in Fig. 6.

\begin{figure}[tbp]
\includegraphics[width=2.8in ]{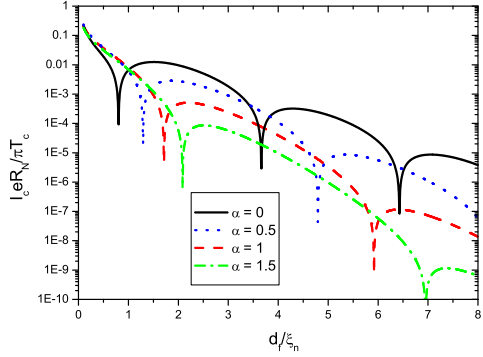}
\caption{(Color on line) Influence of the spin-flip scattering parameter $%
\protect\alpha$ on the thickness dependence of the critical current in S/F/S
junction for $\protect\alpha_{so}=0,$ $h=3\protect\pi T_{c}$, $\protect\gamma%
_{B}=10$, and $\frac{T}{T_{c}}=0.5$.}
\end{figure}

One can see that with increasing $\alpha$, the critical thickness $d_{f1}$\
of the first $0-\pi$ crossover shifts to larger values of $d_{f}$.

\subsection{Influence of spin-orbit scattering on I$_{c}$}

If spin-orbit scattering is not negligible, then the ratio of the
characteristic lengths in the decaying solution (\ref{Id1}), (\ref{Id2})
becomes
\begin{equation}
\frac{\xi_{f1}}{\xi_{f2}}=\frac{\sqrt{1-\alpha_{so}^{2}}}{(\left\vert
\widetilde{\omega}\right\vert +\alpha_{so}+\sqrt{\widetilde{\omega}%
^{2}+2\left\vert \widetilde{\omega}\right\vert \alpha_{so}+1})}.
\label{R_so}
\end{equation}
Eq.(\ref{R_so}) shows that the difference between the decaying $\xi_{f1}$
and the oscillating length $\xi_{f2}$ increases with $\tau_{so}^{-1}$ even
faster than for the case of spin-flip scattering. Moreover, the transition
to monotonically decaying solution takes place at $\tau_{so}^{-1}\rightarrow
h$ and $\xi_{f2}\rightarrow\infty.$

The dependence of critical current on $d_{f}$\ calculated from (\ref{Ic})
for $T=0.5T_{c},$\ $h=3\pi T_{c}$\ and $\gamma_{B}=10$\ is shown in Fig. 7.
One can see that the oscillation period increases strongly with increasing $%
\alpha_{so}$ and diverges when $\alpha_{so}=1$. At the same time, $d_{f1}$\
shifts to larger values of $d_{f}$.

\begin{figure}[tbp]
\includegraphics[width=2.8in ]{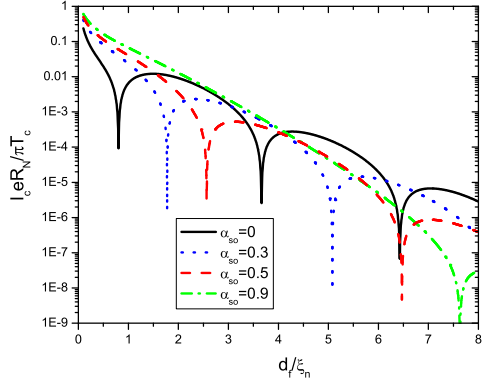}
\caption{(Color on line) Influence of the spin-orbit scattering on the
thickness dependence of the critical current in S/F/S junction for $\protect%
\alpha=0$, $h=3\protect\pi T_{c}$, $\protect\gamma_{B}=10$, and $\frac{T}{%
T_{c}}=0.5$.}
\end{figure}

\section{Critical current of long junctions with transparent interfaces at
arbitrary temperatures.}

Let us now consider a S/F bilayer with a transparent interface. The complete
nonlinear Usadel equation in the F layer has to be employed. For the easy
axis of the ferromagnet and using the usual parametrization of the normal
and anomalous functions $G_{f}=\cos \theta _{f}$ and $F_{f}=\sin \theta _{f}$%
, it may be written in the absence of spin-orbit scattering as%
\begin{equation}
\xi _{n}^{2}\frac{\partial ^{2}}{\partial x^{2}}\theta _{f}-\left( \frac{%
\left\vert \omega \right\vert +ihsign(\omega )}{\pi T_{c}}+\frac{\cos \theta
_{f}}{\pi T_{c}\tau _{m}}\right) \sin \theta _{f}=0,  \label{Usadel_NL_F}
\end{equation}%
Note that Eq.(\ref{Usadel_NL_F}) transforms into (\ref{Usadel_lin_F}) in the
limit of small $\theta _{f}\ll 1.$

For S/F/S junctions, this equation may be used and an analytical solution
found if $d_{f}\gg\xi_{f1}$. In that case, the decay of the Cooper pairs
wave function in first approximation occurs independently near each
interface \cite{buzdinrev}. It can therefore be treated separately enough to
consider the behavior of the anomalous Green's function near each S/F
interface, assuming that the F layer thickness is infinite.

For one interface ($x=-d_{f}/2)$, a first integral of (\ref{Usadel_NL_F})
leads to
\begin{equation}
\xi_{n}\frac{d}{dx}\frac{\theta_{f}}{2}=\sqrt{\frac{\left\vert \omega
\right\vert +ihsign(\omega)}{\pi T_{c}}}\sin\frac{\theta_{f}}{2}\sqrt{\left[
1+\frac{1}{\tau_{m}\left( \left\vert \omega\right\vert +ihsign(\omega
)\right) }\cos^{2}\frac{\theta_{f}}{2}\right] },  \label{FirstI}
\end{equation}
where the boundary conditions $\theta_{f}\left( x\rightarrow\infty\right) =0$
has been used. Further integration in (\ref{FirstI}) gives \cite{courtois}
\begin{equation}
\frac{\sqrt{\left[ 1-\varepsilon^{2}\sin^{2}\frac{\theta_{f}}{2})\right] }%
-\cos\frac{\theta_{f}}{2}}{\sqrt{\left[ 1-\varepsilon^{2}\sin^{2}\frac{%
\theta_{f}}{2})\right] }+\cos\frac{\theta_{f}}{2}}=g_{0}\exp\left\{ -2q\frac{%
(d_{f}/2+x)}{\xi_{f}}\right\}  \label{g0}
\end{equation}
where $k=\sqrt{2}/\xi_{f}\sqrt{(\left\vert \widetilde{\omega}\right\vert
+isign(\omega)+\alpha)}=q/\xi_{f},$ and $\varepsilon^{2}=\alpha/\left(
\left\vert \widetilde{\omega}\right\vert +isign(\omega)+\alpha\right) .$

The integration constant $g_{0}$ in (\ref{g0}) should be determined from the
boundary conditions at S/F interfaces. For simplicity we will assume that
the rigid boundary conditions are valid at $x=-d_{f}/2,$ then
\begin{equation}
\theta_{f}(-d_{f}/2)=\arctan\frac{\left\vert \Delta\right\vert }{\omega}
\label{teta_In}
\end{equation}
and from (\ref{g0}), (\ref{teta_In}) we finally get
\begin{equation*}
g_{0}=\frac{\left( 1-\varepsilon^{2}\right) \mathcal{F}\left( n\right) }{%
\left[ \sqrt{\left( 1-\varepsilon^{2}\right) \mathcal{F}\left( n\right) +1}+1%
\right] ^{2}},
\end{equation*}
\emph{\ }and
\begin{equation*}
\mathcal{F}\left( n\right) =\frac{\left( \Delta/\left( 2\pi T\right) \right)
^{2}}{\left[ n+1/2+\sqrt{\left( n+1/2\right) ^{2}+\left( \Delta/\left( 2\pi
T\right) \right) ^{2}}\right] ^{2}}=\frac{\left\vert \Delta\right\vert ^{2}}{%
\left( \omega+\Omega\right) ^{2}},
\end{equation*}
where $\Omega=\sqrt{\omega^{2}+\left\vert \Delta\right\vert ^{2}}.$ Next, $%
F_{f}$ may be determined.

The anomalous Green's function at the center of the F layer in the S/F/S
junction may be taken as the superposition of the two decaying functions
\cite{likharev} taking into account the phase difference in each
superconducting electrode,
\begin{equation}
\theta_{f}=\frac{4}{\sqrt{1-\varepsilon^{2}}}\sqrt{g_{0}}\left( \exp\left\{
-q\frac{(d_{f}/2+x)}{\xi_{f}}-i\frac{\varphi}{2}\right\} +\exp\left\{ q\frac{%
(x-d_{f}/2)}{\xi_{f}}+i\frac{\varphi}{2}\right\} \right)
\end{equation}
As a result, the current-phase relation is sinusoidal and the critical
current becomes

\begin{equation}
I_{c}R_{N}=64\pi \frac{d_{f}T}{\xi _{f}e}\Re\left( \overset{\infty }{%
\underset{0}{\sum }}\frac{\mathcal{F}\left( n\right) q}{\left[ \sqrt{\left(
1-\epsilon ^{2}\right) \mathcal{F}\left( n\right) +1}+1\right] ^{2}}\exp (-q%
\frac{d_{f}}{\xi _{f}})\right) .  \label{Ic2}
\end{equation}

The critical current is proportional to the small factor $\exp(-qd_{f}/\xi
_{n})$. The terms neglected in our approach are much smaller and are of the
order of $\exp(-2qd_{f}/\xi_{n})$. Therefore, they give a tiny second
harmonic term in the current-phase relation.

It should be underlined that this expression coincides with the one
previously obtained in \cite{kuprianov} in the limit $\tau_{m}^{-1}%
\longrightarrow0$ (no magnetic scattering) and for $T_{c}<<h$.

In addition, note that in the limit of vanishing magnetic scattering, the
temperature dependence of the critical thickness $d_{fk}$\ of F layer when $%
I_{c}=0$ may appear only through the variation of$\ k$ with $\omega $ (where
$k\sim \sqrt{2\left( \left\vert \widetilde{\omega }\right\vert +isign(\omega
)\right) /\xi _{f}}$ in this case). As the characteristic range of the
variation of the Matsubara frequencies in the sum in (\ref{Ic2}) is $\omega
\thicksim \pi T_{c}$ , then in the limit $h\gg \pi T_{c},\tau _{m}^{-1}$,
the dependence of $d_{fk}$ on $T$ will be weak. However, when the spin
scattering is not weak, i.e. $\tau _{m}^{-1}\gtrsim h,$ another mechanism of
the temperature dependence of the $d_{fk}$\ emerges due to the temperature
dependent term $\tau _{m}^{-1}\cos \theta _{f}$ in the Usadel equation (\ref%
{Usadel_NL_F}), or, in another words, due to the complex $\omega $ dependent
function $1-\varepsilon ^{2}.$ For a strong ferromagnet, $h\gg \pi T_{c},$
the latter mechanism may become stronger than the 'thermal' one related to $%
k(\omega ).$ It is not difficult to take into account both mechanisms in the
numerical calculation of the sum (\ref{Ic2}).

As an illustration, we present in Fig. 8 the theoretical fit of the
experimental data for NbCu$_{0.52}$Ni$_{0.48}$Nb junctions by Sellier
\textit{et al.} \cite{sellier2}, making use of Eq.(\ref{Ic2}) valid in the
limit of small interface resistances ($\gamma _{B}=0$). The F layer
thickness used for fit is $d_{f}=18$ nm while the experimental value
presented in \cite{sellier2} is $19$ nm. The difference may be explained by
the uncertainty in F layers determination that is around $1$ nm, and may
even increase due to the presence of a magnetically dead layer near S/F
interface. Taking this in mind, the theoretical description of the critical
current temperature dependence can be considered as rather satisfactory.
Besides, Houzet \textit{et al.} \cite{houzet} have performed numerical
calculations and got also good fit of another experimental curve of \cite%
{sellier2} ($d_{f}=17$ nm) for the set of parameters which are in the same
range that we have used.

\begin{figure}[tbp]
\includegraphics[width=2.8in ]{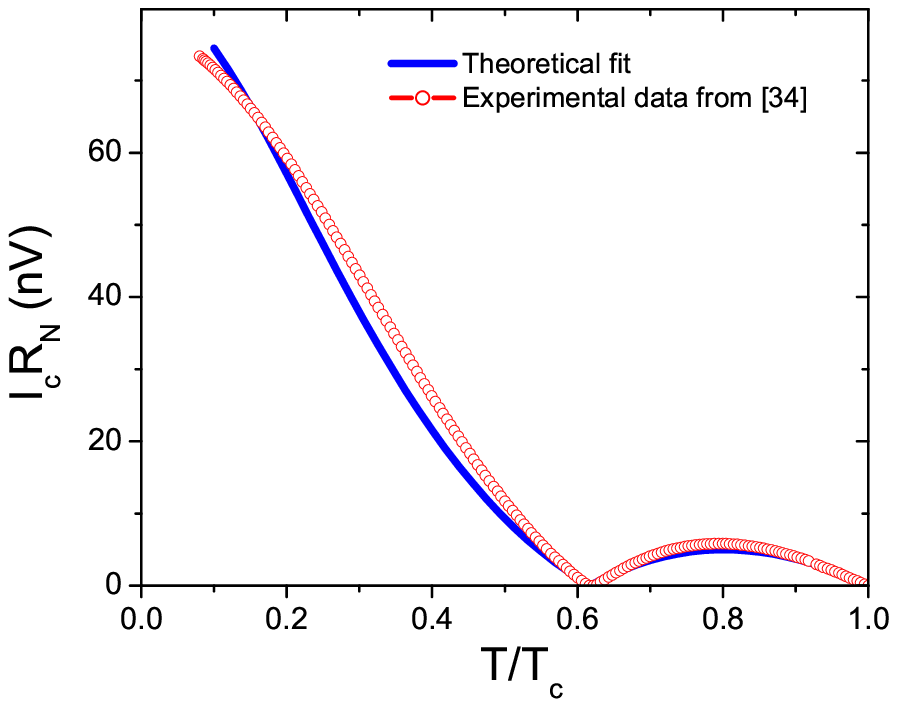}
\caption{(Color on line) Fit to the experimental data from Ref.\protect\cite%
{sellier2} for the critical current in a NbCu$_{0.52}$Ni$_{0.48}$Nb
junction. The fitting parameters are: $h=220K$ and $\protect\alpha
=3$.}
\end{figure}

As another application of the formalism, we present in Fig. 9 the
theoretical fit of the experimental data of Ryazanov \textit{et al.} \cite%
{RyazanExp} for NbCu$_{0.47}$Ni$_{0.53}$Nb junctions with $d_{f}=22$ nm.\
Good agreement is achieved assuming $h=650$ $K$, $\alpha =1.35$ and $%
d_{f}=21 $ nm. As in the previous fit, small difference in $d_{f}$ may be
explained by the uncertainty in F layers determination.

\begin{figure}[tbp]
\includegraphics[width=2.8in ]{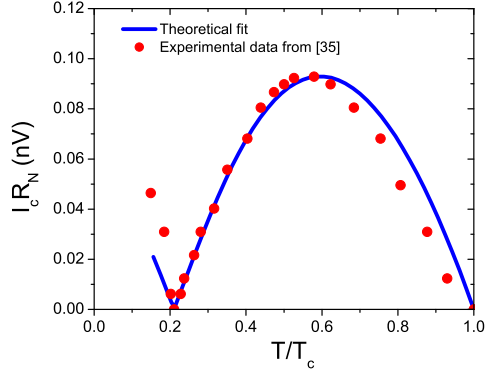}
\caption{(Color on line) Fit to the experimental data from Ref.\protect\cite%
{RyazanExp} for the critical current in a NbCu$_{0.47}$Ni$_{0.53}$Nb
junction. The fitting parameters are: $h=650K$ and $\protect\alpha
=1.35$.}
\end{figure}

\bigskip

Note that the rather complicated expression for the critical current (\ref%
{Ic2}) simplifies near $T_{c}$ and may be written as (for $T_{c}<<h$)

\begin{equation}
I_{c}R_{N}=\frac{\Delta ^{2}\pi d_{f}}{2eT_{c}\xi _{f2}}\frac{1}{\cos \left(
\Psi \right) }\exp (-\frac{d_{f}}{\xi _{f1}})\sin \left( \Psi +\frac{d_{f}}{%
\xi _{f2}}\right) ,
\end{equation}%
where\ $\Psi $ is defined by $\tan (\Psi )=\dfrac{\xi _{f2}}{\xi _{f1}}$.
The damping oscillatory behavior of the critical current is therefore
retrieved, and the simplicity of the previous expression makes it useful for
theoretical description of the evolution of the critical current with the
thickness of the F layer.

\bigskip

\section{Conclusion}

We have made detailed theoretical investigation of the influence
spin-dependent scattering in the ferromagnet on the critical temperature of
S/F multilayered systems and the critical current of S/F/S Josephson
junctions. More precisely, we have demonstrated that spin-flip and
spin-orbit scatterings both lead to the decrease of the decaying length and
the increase of the oscillations period. Besides, spin-orbit scattering may
be more harmful for superconductivity than spin-flip scattering. Indeed, the
oscillations of $T_{c}$ and $I_{c}$ can be destroyed by spin-orbit
scattering, while spin-flip scattering can only modify them. This allows to
distinguish these pair-breaking mechanisms, which both should be taken into
account for theoretical fits of experimental data. Moreover, the
simultaneous introduction of the spin-flip and the spin-orbit scattering
leads to important new prediction that the perpendicular spin flip can
compensate spin-orbit and vice versa. Then the perpendicular magnetic
scattering may restore the oscillations of $T_{c}$ and $I_{c}$ which would
be otherwise absent if spin-orbit scattering is strong enough.

We have also studied the influence of the interface transparency on $T_{c}$
and $I_{c}$ in S/F and S/F/S structures, respectively. The nonmonotonous
behavior of $T_{c}$ with the interface transparency parameter $\gamma _{B}$
was predicted in the case of thin F-layers due to multiple scattering at the
interfaces. We have shown that the same mechanism is responsible for the 0-$%
\pi $ transition in S/F/S junctions with small thickness $d_{f}$ in the case
of large $\gamma _{B}$. For larger $d_{f}$ detailed analytical and numerical
study of the influence of $\gamma _{B}$ parameter is presented.

It was predicted in Ref.\cite{bverev} that in the case of noncollinear
magnetic ordering the long-ranged triplet component of the superconducting
condensate could be generated. The spin-orbit and the perpendicular magnetic
scattering are rather harmful for this long-ranged triplet component. Our
analysis shows that the ferromagnetic alloys like Cu$_{x}$Ni$_{1-x}$, where
following our estimates the parameter $1/h\tau _{m}$ may exceed unity, are
not suitable candidates for the experimental search for the triplet
component.

Finally, note that it may be interesting to study the Josephson junctions
with the ferromagnetic layer substituted by a paramagnetic one. Applying an
external magnetic field $H$, it is possible to produce the polarization of
the magnetic atoms and then generate an internal field $h\sim \chi H$, where
$\chi $ is the paramagnetic susceptibility. The variation the external field
allows to change the relative contribution of the scattering mechanisms and
the exchange field and then to modify the properties of the junction in a
controllable way. In particular, it could provoke the transition from 0 to $%
\pi $ state.

The authors thank M. Daumens, M. Houzet, M. Kulic, C. Meyers, and
V.V. Ryazanov for useful discussions. We also thank H. Sellier and
V.V. Ryazanov for communication of their experimental data. This
work was supported in part by EGIDE Programme 10197RC, ESF PI-Shift
Programme, RFBR grant N 06-02-90865 and NanoNed programme under
project TCS.7029.

\section{APPENDIX:}

The coefficients $A_{\omega\pm}$ and $B_{\omega\pm}$ that appear in (\ref%
{SolP}) and (\ref{SolM}) may be written as

\begin{equation}
A_{\omega+}=\frac{2\Delta G_{S}\cos(\varphi/2)}{\left\vert \omega\right\vert
\left\{ \gamma_{B}\xi_{n}k_{+}\sinh\left[ k_{+}d_{f}\right] +\left\vert
G_{S}\right\vert \cosh\left[ k_{+}d_{f}\right] \right\} (1+\eta_{\omega
}^{2})},
\end{equation}%
\begin{equation}
B_{\omega+}=\frac{2i\Delta G_{S}\sin(\varphi/2)}{\left\vert
\omega\right\vert \left\{ \gamma_{B}\xi_{n}k_{+}\cosh\left[ k_{+}d_{f}\right]
+\left\vert G_{S}\right\vert \sinh\left[ k_{+}d_{f}\right] \right\}
(1+\eta_{\omega }^{2})},
\end{equation}%
\begin{equation}
A_{\omega-}=-\frac{2\Delta G_{S}\eta_{\omega}\cos(\varphi/2)}{\left\vert
\omega\right\vert \left\{ \gamma_{B}\xi_{n}k_{-}\sinh\left[ k_{-}d_{f}\right]
+\left\vert G_{S}\right\vert \cosh\left[ k_{-}d_{f}\right] \right\}
(1+\eta_{\omega}^{2})},
\end{equation}%
\begin{equation}
B_{\omega-}=-\frac{2i\Delta G_{S}\eta_{\omega}\sin(\varphi/2)}{\left\vert
\omega\right\vert \left\{ \gamma_{B}\xi_{n}k_{-}\cosh\left[ k_{-}d_{f}\right]
+\left\vert G_{S}\right\vert \sinh\left[ k_{-}d_{f}\right] \right\}
(1+\eta_{\omega}^{2})},
\end{equation}
The symmetry relations following from (\ref{SolP}) -(\ref{EtiBe}) are at $%
h\leq\tau_{so}^{-1}$
\begin{equation}
\eta_{\omega}=\eta_{-\omega}^{\ast},\quad A_{-\omega+}^{\ast}=A_{\omega
+},\quad A_{-\omega-}^{\ast}=A_{\omega-},\quad
B_{-\omega+}^{\ast}=-B_{\omega+},\quad B_{-\omega-}^{\ast}=-B_{\omega-}.
\end{equation}
and if $h\geq\tau_{so}^{-1}$:
\begin{equation}
\begin{array}{c}
\eta_{-\omega}^{\ast}\eta_{\omega}=-1,\quad\beta_{\omega+}=\beta_{\omega
-}^{\ast}=\beta, \\
\eta_{-\omega}^{\ast}A_{-\omega+}^{\ast}=A_{\omega-},\quad
A_{-\omega-}^{\ast
}=\eta_{\omega}A_{\omega+},\quad\eta_{-\omega}^{\ast}B_{-\omega+}^{\ast
}=-B_{\omega-},\quad B_{-\omega-}^{\ast}=-\eta_{\omega}B_{\omega+}.%
\end{array}%
\end{equation}
These relations allow us to simplify the calculations for the determination
of the critical current.

\end{document}